\documentclass[letterpaper]{article} 
\usepackage{aaai23}  
\usepackage{times}  
\usepackage{helvet}  
\usepackage{courier}  
\usepackage[hyphens]{url}  
\usepackage{graphicx} 
\usepackage{natbib}  
\usepackage{caption} 
\title{One Artist's Personal Reflections on Methods and Ethics of Creating Mixed Media Artificial Intelligence Art}
\author{
    Jane Adams, M.F.A.
}
\affiliations{
    Khoury College of Computer Sciences at Northeastern University\\

    440 Huntington Ave, Rm. 302\\
    Boston, Massachusetts 02115 USA\\
    jane@universalities.com
}

\begin{document}

\maketitle

\begin{abstract}
I intend to make a scientific contribution of my subjective experience as a single unit of self-described ``artist" leveraging artificial intelligence as an assistive visual creation tool, in the hopes that it may provide some inspiration or deeper meaning for fellow artists and computer scientists in this medium. First, I will provide some background on my personal history thus far as an artist. Neither artist nor scientist can exist in a vaccuum, so I then will provide some (albeit a non-exhaustive list of) related work that has helped me contextualize my own work and thinking in this area. I often consider my methods in the creative process chronologically, so I have divided that section according to the loose structure of my artistic workflow. These foundations provide a fertile grounding for discussion around topics of subject matter, reception, community, and ethics. I then conclude with some ideas for future work in the realms of theory of authorship, explainability tooling, and research framing.
\end{abstract}

\section{Introduction}
From December 2020 through September 2022, I embarked on a journey in creating art using machine learning. An emergent media artist by study and practice, I found myself engaging in a metalogue about the ethical and artistic implications and considerations of my work, concurrent to my actual creating. Upon reflexively pausing my work with this medium after the summer of 2022 and the explosion of text-to-image models, I've begun a process of cataloguing my experience and creating reflective artifacts. In particular, here I hope to share some relevant background related to the history of mechanized art creation; philosophies on the role of `artist' in the era of remix and collage; and the use of artificial intelligence as a creative tool. Following this grounding, I review my own methods for data curation and pre-processing; model development and iteration; and mixed media output. I discuss the subject matter and reception of my work; experiences in building impromptu community around these topics; and ethical considerations of artificial intelligence artwork including: provenance, identity of authorship, intellectual property, class, and privacy. In future work, I consider these discussions with an eye toward the emergent landscape of machine learning and creativity.

\subsection{Background}
With a background in graphic design and digital media, much of my early artistic work in my teens was centered around collage. I was particularly drawn to digital collage work, such as Mark Weaver's collection \textit{Make Something Cool Every Day} \cite{Weaver2009MakeSomething} and magazines like \textit{NYLON} and \textit{i-D}. In my undergraduate studies, I became familiar with interactivity as a new modality for creating artistic experiences, and delighted in using tools like \textit{Processing} \cite{Reas2006Processing} to replicate the works of famous generative artists, e.g. creating a code version of Sol Lewitt's \textit{Wall Drawing {\char"0023}295} \cite{Lewitt1976Walldrawing}.

As a graduate student studying emergent media, I became engrossed in the possibilities of a physical studio practice, swept up in the new wave of ``maker spaces" to experiment with physical computing (e.g. Arduino and Raspberry Pi), woodworking, welding, laser cutting, and 3D printing. Transmedia approaches fascinated me: the notion that you could take a photograph, edit it on a computer, laser engrave it onto a piece of wood, roll that wood across ceramic and kiln fire it, then photograph that sculpture and render it in virtual reality -- the barrier between the worlds of physical and digital had never been so permeable \cite{jenkins2006transmedia,Jenkins2004MediaConvergence}.

I began to think about artistic subject matter as data, with an interesting dichotomy: that ``the medium is the message" in the words of Marshall McLuhan \cite{McLuhan1967medium}, but that also ``The fundamental problem of communication is that of reproducing at one point either exactly \textit{or approximately}* a message selected at another point", in the words of information theorist Claude Shannon \cite{shannon1948mathematical} \textit{(* emphasis added)}. The ``approximately" part was the part that interested me most as an artist; how data is transformed and distorted in its journey across media. It was this convergence of arts and data that drew me to professional work in data visualization.

While much of the artistic media I created in 2020 was large-scale, sculptural, and interactive, I was professionally working adjacent to researchers in complex systems and machine learning, and the creative potential of their tools intrigued me. Helena Sarin's 2019 talk "Playing a Game of GANstruction" at EyeO Festival, which I attended in person, was deeply inspiring \cite{Sarin2019GANstruction}. However, machine learning was an intimidating medium, and I had mostly remained awestruck on the sidelines. It took a global pandemic to stir me to action.

As the world shut down, so too did the art gallery that housed an aquaponic sculpture I had built: a fully plumbed and electrified dollhouse exploding with tiny plants, supported below by a homeostatic micro-environment of fish and algae and snails. This was not the kind of work to continue making while quarantined in a small apartment. So began my adventure into artificial intelligence tools as creative assets, which presented a whole world of complex phenomena and interesting possibility, with the benefit of being able to be tucked away in a laptop beneath a growing mound of books and entertainment ephemera.

\section{Related Work}
There are many works that have informed my adventures in creating art with the help of machine learning technologies; I regret that I had not been more diligent in documenting them during my process. However, in reflecting, I find that three categories of art in particular feel foundational: mechanized art, remix and collage art, and artificial intelligence art.

\subsection{Mechanized Art}
The history of mechanized production of art predates the computer, the jacquard loom, the printing press, the woodblock print, the wax seal, and the stone tablet \cite{gombrich1995story}. However, with each new method of mechanized production comes new discourse on the meanings of `creativity' and `artist'. Consider the advent of photography, during which ``because of the alleged essential differences between photographs and other kinds of pictures, there is a total discontinuity between the history of the manipulative graphic arts and the pre- and early history of photography” \cite{snyder1982galassi}. This early perspective on photography as `other" from all prior graphic arts has since been resoundingly overturned to support the contemporary view that ``Photographs and, say, paintings, represent in the same way and for the same reasons”\cite{snyder1982galassi, gombrich1995story}. One is not surprised to find the work of photographer Ansel Adams in an art gallery today, but seeing artists who use newer mechanized technologies like artificial intelligence in a gallery might come as a shock to some who would decry it ``not art" just like the prescriptivists of the 1850s.

It may be useful here also to note the difference Steven Jones identifies in \textit{Against Technology}: between the Luddites, who were not anti-technologist but rather acting as organized laborers protesting an economic injustice; and Neo-Luddism, a more modern romantic ideal of simple living in the face of ever more technology \cite{jones2013against}. With any new technology, we must be astute in making similar distinctions among its detractors.

The philosopher Walter Benjamin wrote in his seminal 1935 essay ``The Work of Art in the Age of Mechanical Reproduction" about the \textit{``aura"} of a work, which he argues is diminished with each subsequent reproduction \cite{benjamin2018work}. It is this framing, of diminishing originality, that explains too the economics of selling a print of a painting for less than the painting itself. How Benjamin's conceptualization of the aura fits into discussions of transformative use and the dilution or imitation of training data remains an interesting discussion in the era of machine learning.

From 2009 to 2016, a webzine entitled ``Notes on Metamodernism" collected discourse from critics and theorists across disciplines and around the world in an attempt to characterize 21st century culture in the wake of postmodernism \cite{vermeulen2010metamodernism}. While postmodernism, arguably anticipated by Dadaism, was known for deconstructionist views, atomizing the very nature of knowledge and being, metamodernism offers a return to context through the understanding of mutability \cite{storm2021metamodernism}. For example, the metamodernist manifesto begins, "1. We recognise oscillation to be the natural order of the world", and one of these such oscillations is that ``Today, we are nostalgists as much as we are futurists " \cite{turner2015manifesto}. I posit that much of the artificial intelligence art being made today is of a quintessentially metamodernist nature, though perhaps not cognizantly so. 

I would be remiss were I not to mention McKenzie Wark's 2004 \textit{Hacker Manifesto}, which introduces the conception of the ``hacker" class and the ``vectoralist" class in a critique of information commodification \cite{wark2004hacker}. Wark's intentionally broad definition of ``hacking", coupled with her pointed opposition to vectoralist appropriation of previously free information, is deeply relevant to discussions today about the provenance and authorship of artificial intelligence art.

\subsection{Remix and Collage Art}
A natural complement to mechanized reproduction of art, and with an equally lengthy history, is its remix and collage into something new, a process of disassembly and synthesis \cite{frosio2021brief}. The advent of computers ushered in a new era of remix, and the further expansion of remix beyond music to image and video synthesis through software such as Adobe Photoshop, Windows Movie Maker, and iMovie 
\cite{Knobel2008Remix}. Along with remix culture came new musings and grievances in the realm of intellectual property and copyright law \cite{lessig2004free}. It is possible actually that because remix in music predated slightly the era of remix in image and video media, one might look toward stories like that of Danger Mouse's controversial \textit{Grey Album} -- which spurred reactionary academic activism against the growing reach of intellectual property law -- as fables by which to draw parallels to the current discourse. This academic activism in turn cites earlier, non-computational images produced by mechanized processes, such as collage, in its decrying of copyright overreach \cite{kembrew2005dangermouse}.

One might consider even the fraught history of pop artists like Andy Warhol as a useful frame of reference for earlier discussions on remix and collage. Famously, the Andy Warhol Foundation is embattled in an ongoing copyright lawsuit with photographer Lynn Goldsmith, presently in the hands of the U.S. Supreme Court, over a 1984 silkscreen print of the musician Prince  \cite{epstein2022sequential}. The outcome of this decision on what constitutes `fair use' and `transformative use' will likely ripple across much of the art world and the way copyright is enforced in the future \cite{npr2022prince}.

In 2012, Austin Kleon published his tips for creatives provocatively entitled "Steal Like an Artist". In it, Kleon unabashedly advocates for theft of ideas, because ``What a good artist understands is that nothing comes from nowhere. All creative work builds on what came before. Nothing is completely original." What he is not advocating is to    `rip off', but rather what we have been discussing: to remix, collage, transform \cite{kleon2012steal}. This section of his book I find particularly liberating and yet equally cautionary, tempering: he advocates for `theft' as a shorthand for labored obfuscation, the laundering of ideas, as contrasted with \textit{copying}, or direct imitation. In a way, he has conveyed the spirit of `fair use' law more loudly and simply, in the way only an artist can. Notably, Lawrence Lessig chronicles how, if modern copyright law were applied to the cultural artifacts of even the past century, much of modern U.S. art would not have been legally created \cite{lessig2004free}.

\subsection{Artificial Intelligence Art}
While the history of artificial intelligence art is short in years, it is incredibly dense in innovation, with NVIDIA CEO Jensen Huang in 2017 describing the rapid development of new machine learning models as a ``Cambrian Explosion", in reference to the `biological big bang' of abrupt, immense species diversification of life on earth some 541 - 485.4 million years ago \cite{Leopold2017Cambrian}.

The 2021 survey paper \textit{Understanding and Creating Art with AI: Review and Outlook} identifies the major technological milestones in this area as: the introduction of Generative Adversarial Networks (GANs) in 2014 \cite{mirza2014conditional}; DeepDream in 2015 \cite{mordvintsev2015inceptionism}; Neural Style Transfer (NST) in 2016 \cite{gatys2016image}; AICAN in 2017 \cite{elgammal2017can}; and DALL-E \cite{reddy2021dall} and CLIP \cite{radford2021learning} in 2021 \cite{Cetinic2021Survey}. Even since then, text-to-image has taken off, including in 2022 alone: DALL-E 2 \cite{ramesh2022dalle2}, Imagen \cite{imagen2022}, MidJourney \cite{midjourney22}, etc. and the advent of text-to-video \cite{singer2022textvideo} and text-to-3D \cite{poole2022dreamfusion} models.
The 2022 survey \textit{GAN computers generate arts?} categorizes image-based GANs into four sections: Conditional GANs, DCGAN, Standard GAN, and Style-Based GAN; and details the tasks, loss function(s), architecture, and results of papers in each section \cite{Shahriar2022GANSurvey}. I will admit that as an artist with limited formal training in computer science, the alphabet soup of these models has been a deep pool, but I have found some through my practice that became particularly near to my heart, as I will elucidate in the next section.

In terms of artists and artworks that have particularly moved and inspired me, I lament page restrictions, because there are so many to list. However, I will mention a few here. First, Helena Sarin's \textit{Leaves of Manifold} inspired me to collect my own training data \cite{helena2018leaves}. The \textit{Neural Zoo} by Sofia Crespo reminded me of my teenage love for collage, and showed how similar strategies could be applied to the products of GANs \cite{crespo2022neural}. I also loved the Style Transfer works of Matthew Seremet and the artist known only as luluixixix, not only for the aesthetic qualities of their work but for the humble, populist personas they carried as artists \cite{Seremet2021spaghetti, lulu2019}.

The controversial sale of the \textit{Edmond de Belamy} by the artist collective Obvious was fascinating to me in my considerations on authorship, because the model that the image came from had been trained by artist-developer Robbie Barrat, who was not compensated from the \$432,500 auction at Christie's \cite{gillotte2019copyright, miller2019isitart}. Similarly fascinating from an authorship perspective was the touting of art `created' by humanoid robot Sophia in 2021 which sold for nearly \$700,000 \cite{karimova2020adaptation, parviainen2021political}. I found it not insignificant that both of these stories of questionable authorship entered public discourse not only for their philosophical conundrums, but also in part because of the vast sums of money involved. Many of these nuanced moral questions of attribution seem invariably to fall more concretely over time into arguments over compensation. In the first instance I felt that Barrat should have been compensated for his development of the model, primarily because the Obvious collective hadn't done any subsequent training of their own; whereas in the second instance, the anthropomorphization of the robot struck me as a trite marketing strategy that undermined the contributions of the humans responsible for the robot's development and the curation of the art produced. Perhaps I too am a Neo-Luddite in my anthropocentrist beliefs on this matter, to deny the androids their aura.

Stylistically, I was drawn to StyleGAN2 as a model for its distinct `visual indeterminacy', as described by Aaron Hertzmann in relation to A.I. art \cite{hertzmann2020visualIndeterminacy}. The confounding nature of indeterminant images is, I think, profoundly metamodernist. But on a more simple level, there is something less uncanny and more comforting about the `uncanny valley' of 2020-2022 A.I. art: specifically, the ability to still possibly differentiate between what is human-made and what is machine generated. Every time a fragment of mountain is rendered in a cloud, or a sixth finger emerges from a handshake, or an iconic fast food sign is given a makeover in indecipherable hieroglyphs, we might breathe another sigh of relief that there are still CAPTCHAs yet unsolvable by machines.

\section{Methods}
I roughly categorize my creative process into three ordered steps: pre-training, training, and post-training. This comes from my study of Norbert Weiner's Cybernetics, and my conceptualization of models as 3-component systems comprised of stimulus (input), message (transformation), and response (output) \cite{wiener1954cybernetics}.

\subsection{Data Curation and Pre-Processing}
The majority of my work involved training StyleGAN2 models using collected data through a combination of manual and automated means. Some datasets were collected by thumbing through pages of royalty-free stock photo websites or public image archives and building `collections' directly through the site's user interface, a process which would take between 5 and 10 hours to build a collection of 2,000 images. Many of these datasets were then artificially inflated in size through cropping, rotation and mirroring, or other perturbations such as hue shifting. Other datasets were collected through more automated means, using scripts like BeautifulSoup or emulators like Selenium to bypass pagination and traverse whole sites or branches of site architectures.


Likewise, sorting and filtering of datasets was done through a combination of manual and automated means. I had to replace the `down' key on my laptop keyboard when it came loose after scanning through thousands of images to ensure they met my holistic criteria for the model I intended to build. My laptop also suffered from overheating (I once documented it at 107$^{\circ}$F / 41.7$^{\circ}$C using a laser thermometer) during more automated filtration methods. One such method was k-means clustering; I had used FFMPEG to extract video frames from a collection of animated movies by my favorite animator Hayao Miyazaki, and wanted to filter down to only the landscapes. As a rough approximation of this filter, for each image, I used k-means to cluster pixels into 5 groups, then sorted down to images with pixels in an HSV range within the mid to light blues and greens.

Another dataset bolstering approach I used involved creating a plant-pot adjacency matrix. I had scraped hundreds of potted plants from product photography of home improvement and online shopping storefronts, but my dataset was still too small (in part because plants are so diverse in their morphology) to produce the results I had hoped for in my model. So to artificially increase my training data size, I used Adobe Photoshop and Illustrator to cut out plants and pots respectively, and then pair every permutation of pot and plant with one another, increasing the size of my data to $N^2$.

Photoshop's batch actions proved immensely useful for thinking through scripting actions with the aid of a user interface for graphical reasoning. I used batch actions to develop many of my own rulesets; for one dataset, I created a series of actions to select and mask fashion models off the runway, clip the area surrounding them so all heads would be in approximately the same location, and apply a green screen in the background. The alignment produced a much better output result, and through the use of chroma keying in Adobe After Effects, I was able to then mask the animated people from the latent walk video and create non-square animations.

I also leveraged a number of neural style transfer (NST)-like effects. For example, with High-Resolution Daytime Translation (HiDT), I could increase the size of training data by generating additional images at sunset or night-time \cite{hidt2020}. Similarly, models like CycleGAN made it possible to change the weather in datasets, so if I was training an autumn model, I wasn't restricted only to collecting images of landscapes during the autumn months \cite{CycleGAN2017}. These methods are documented further in my Spring 2021 NVIDIA GTC talk, ``Leveraging Machine Learning in Artistic Workflows" \cite{adams2022nvidia}.

\subsection{Model Development and Iteration}
When I first started training models, I was using RunwayML, an online web interface for model training and exploration. However, that platform quickly became cost-prohibitive, and limited in the features I was eager to use. Knowing Python and having used Jupyter Notebooks for data science and visualization previously, I began experimenting with notebooks in Google Colaboratory (``Colab"), and found them surprisingly easy to use, as well as modular. Some hacks were necessary: I adjusted my sleep cycle slightly so that my creative time aligned with the wee hours of the morning when I was more likely to get an A100 or V100 GPU allocation; someone in one of my A.I. artist Discord channels shared a `keepalive' script to ensure sessions wouldn't time out during training; and I found that keeping files on my own server via sFTP and using \textsc{wget} to transfer back and forth from my Google Drive was much more efficient than directly uploading or downloading from Drive.

I quickly established `genealogies' of transfer learning from one model to another; once I had a model that I particularly liked, it became a new `parent' node for subsequent branching models of different subjects. For example, I trained a model on stock photography of dewy plants, taken with a macro lens. The model became so good at replicating water droplets, that I used it to train a new model on raindrops on glass, and another on a combination of bubbles, candy, and glitter. This `survival of the fittest' of models echoed the earlier design paradigm of Joel Simon's \textit{Artbreeder}, a community-driven art creation site which allows users to collaborate on StyleGAN and BigGan driven A.I. art \cite{simon2020artbreeder, kocasari2021exploring}.

I became interested in stringing together one model after another after seeing how Mikael Alafriz' \textit{LucidSonicDreams} package used .pkl files to apply pre-trained StyleGAN2 weights to generate audioreactive music videos with adjustable parameters \cite{Alafriz2022Lucid}. With the widespread introduction of text-to-image diffusion models, more opportunities arose for daisychaining models together. I used DALL-E 2 \cite{ramesh2022dalle2} to generate stacks of isometric houses to then train a StyleGAN2 model to animate between buildings. In a collaborative project entitled \textit{Mythologicals}, project co-founders Jeremy Torman and Brad Tucker trained a StyleGAN model on images of Pokémon, then used the outputs of those models as a base upon which project collaborators could imagine and iterate on new mythological creatures with magical combinatorial traits \cite{torman2022mythologicals}.

In collaborating with musician Alexa Woodward to create a series of latent walk music videos for her EP \textit{Voyager}, datasets were iteratively built up and honed down to adjust the model output to her vision. For example, for the video ``Waiting on You", we went through four model training cycles: 1) cathedrals and churches from an architectural archive, initially selected because of the religious conntations in the song; 2) forest images from an image classification repository, as we pivoted to a more nature-inspired representation of spiritual connectedness; 3) a second forest image set collected manually from stock photo websites after the second dataset was too grainy and low resolution; and finally 4) an expansion of the third dataset featuring forests, but now with the addition of wide open spaces like lakes and deserts inspired by the musician's memories of childhood in the desert. The final result was a grand tour of undulating landscapes, contracting into forests and expanding out to the horizon, like a time lapse animation on a geologic scale.


\subsection{Output Media}

Continuing my multimedia explorations, I hoped to break out of the confines of the 1024x1024 pixel box that was the default output of StyleGAN models. I created digital collages using Photoshop, then taught myself advanced strategies in Adobe After Effects to create surreal cinemagraphs: illusions of melting maps, moving photographs, and shifting landscapes framed by windows and cameras.

Hans Ulrich Obrist is a curator who has written extensively about the tragedy of the iconic `white cube' art gallery, lamenting the way such a sterile framing divorces art from artist \cite{obrist2014ways}. In the confines of my home, I wanted desperately to again visit galleries that had been curated to showcase art and artist together; to attend talks and wander hallways and stare up at works larger than my own being. I found this opportunity in the form of galleries in the `metaverse': bizarre places like Decentraland \cite{Decentraland} and Voxels \cite{Voxels}. While these platforms provided visitors and artists alike anonymity by virtue of being virtual worlds, they were by no means white cubes; they abounded with color and animation, music blared from speakers and brightly decorated avatars apparated and disapparated from the 3D realm at random. I participated in several group art shows in these spaces, including one show in Decentraland specifically for women artists located at a virtual skate park, and another in Voxels in a three-story building dedicated entirely to A.I. artists.

As the pandemic subsided and I began to venture once more out into the world, I became particularly curious about how to also free my art from the confines of my screen. Physicalization of my digital work was top of mind. I experimented by generating images with StyleGAN models I had trained, applying subtle DeepDream \cite{mordvintsev2015inceptionism} effects over them, and printing the resulting images onto fabric. I chose fabrics that matched the aesthetic qualities of the images: soft and fuzzy minky fabric for the fluffy clouds of star systems; ethereal and translucent chiffon for elusive whisps of flame; sleek satin for elegant moss-covered forests. Machine learning also came in handy for creating infinitely tilable textures \cite{tilable2022}.


I found ESRGAN, which was useful for upscaling my images for resolutions higher than 1024x1024 \cite{wang2018esrgan}. This enabled me to scale output images large enough to get high-resolution prints made. I had all manner of works printed on paper, canvas, and metal sheets. Applying ESRGAN generated unexpected results in many cases; one model that I had trained on close-up photographs of human bodies, when upscaled, was suddenly covered in a fine layer of `peach fuzz' hair, presumably because the colors matched those of humans in the ESRGAN training data.

I created videos and projection mapped them into my space, making windows appear in solid walls with changing landscapes behind them. I created music videos by manually selecting frames to linearly interpolate between in latent walks \cite{alexa2021waiting}. After learning 3D modeling to mock up the concept, I printed out hundreds of frames of latent walk videos onto overhead projector transparency film, cut them into small squares, and glued them together into a translucent ``latent cube" to imagine a physical form for a vector through high-dimensional space.

\section{Discussion}
\subsection{Subject Matter}
The content of my work often involves patterns of emergence and synergy. In particular, I enjoy the natural phenomena of subjects like trees, mycorrhizal networks, astronomical formations, Lichtenberg figures, and hydrological behavior. The similarities between these phenomena and other concepts that we may think of superficially as uniquely human (data structures, mathematics, logic circuits) are actually so closely intertwined as to be nearly indistinguishable. In choosing this subject matter, however, I am inevitably conflicted then about the environmental impact of using these technologies. From silicon and lithium extraction, to burning of fossil fuels to cool hot server farms for ever more convolutions and renders, the deceptively named `cloud' is alarmingly corporeal \cite{bender2021parrots}. Finally, the subject matter of my work is also self-reflective, in that I often use collage media to think through notions of identity and belonging, including reflections on gender and community.

\subsection{Reception}
I've found that there is profound educational benefit to creating art using any emergent technology, as art has the ability to enchant and yet simultaneously demystify. I found myself torn between wanting to personify ``the A.I.", e.g. ``two neural networks are conversing; one is making the art and the other is judging" versus avoiding any anthropomorphization and flatly declaring it to be ``just statistics all the way down". I've resolved to imagine these two explanatory approaches at either end of a spectrum which it might be beneficial to traverse up and down throughout one's exposition.

\subsection{Community}
Throughout this exploration, I found much more community than expected. From Twitter threads; to Colab notebooks (both freely shared and limited to Patreon supporters); to channels on Discord (like EleutherAI) and Slack (like RunwayML); and communities like r/MediaSynthesis on Reddit; I found deep, scintillating discussions with people from around the world with a wide range of backgrounds, united by our interest in this new technology.

\subsection{Ethics}
I mentioned previously my concerns about the environmental impacts of such models; I think we should not lose sight of the physicality of the cloud, and I became wary too, during my training on Google Colab, that there are unanswered questions about solutions like carbon recapture technology, as detailed in ProPublica's investigative reporting on Carbon Offset Credits \cite{propublica}.

Regarding training data, from my own explorations, I have 
found immensely helpful guidance in the \textit{Responsible AI Field Guide} by the Partnership on A.I. \cite{leibowicz2021fieldguide}. Co-authored by artists, it leaves many questions unanswered, but instead asks individual artists to make their own considerations about the provenance of the data and code they use in their models. In my opinion, the morality of claiming supremacy of copyright over creative freedom is context-dependent;
Matthew Bunker describes in \textit{Eroding Fair Use: The "Transformative" Use Doctrine After Campbell} how the legal definitions and justifications surrounding corporate copyright have begun to shift in recent years towards a more favorable policy for corporations, by requiring the benchmark of `transformative use', and away from creative freedom \cite{bunker2002erodingfair}. This echoes earlier discussions of McKenzie Wark's classification of the hacker vs. vectoralist class \cite{wark2004hacker}; collage as analogy \cite{kembrew2005dangermouse}; and raises questions about the requirements as applied to multimodal A.I. art, eloquently detailed in a particularly interesting case of painted GAN art in Jason Bailey's 2019 article "Why Is AI Art Copyright So Complicated?" \cite{bailey2019copyright}.

\section{Future Work}
\subsection{Authorship}
The 2018 ``Moral Machines" experiment asked participants, in an extended version of the classic `trolley problem' of ethics discussions, to decide how they would like a self-driving car to act in such a scenario \cite{awad2018moral}. I propose a similar study for questions of A.I. art ethics. If an A.I. artist scrapes another living painter's social media account with the intention of copying their style, have they violated Austin Kleon's principle to ``steal \textit{like an artist}"? If I have stolen the product photography of the world's largest online retailer in order to generate new alien plants, should I owe fractional cents to them for every image I have scraped? Only by considering A.I. as a next step in pictorial representation, just as we have evolved to consider photography as such in relation to all prior art, can we possibly place these new moral qualms on a continuum with its peers in other, older, media \cite{snyder1982galassi}. 

\subsection{Explainability}
I am excited by the prospect of explainable A.I. research in other fields being applied to art; both for solving the sticky questions of transformative use and data provenance; and for the simple joys of manipulating model output through the use of a user interface. I particularly enjoyed a phase shift during which Colab notebooks began adding slides and dropdowns for customization, as much for the usefulness of the buttons and sliders themselves as for the care it demonstrated that the community had for sharing and helping newcomers with adoption.

\subsection{Destinations}
What I see as most urgent in an art research context is the necessity to codify language and develop robust taxonomies for discussing these new technologies. Surveys in this area are beginning to surface, but we face a challenge in that the rapid development cycle also means it is hard to stay on top of reading literature, so we ought to be careful not to reduplicate efforts or create conflicting terminologies.

\bibliography{aaai23}

\end{document}